# Speaker Verification in Emotional Talking Environments based on Third-Order Circular Suprasegmental Hidden Markov Model

Ismail Shahin[1], Ali Bou Nassif[2]
Department of Electrical and Computer Engineering
University of Sharjah
Sharjah, United Arab Emirates
[1]ismail@sharjah.ac.ae, [2]anassif@sharjah.ac.ae

*Abstract*—Speaker verification accuracy in emotional talking environments is not high as it is in neutral ones. This work aims at accepting or rejecting the claimed speaker using his/her voice in emotional environments based on the "Third-Order Circular Suprasegmental Hidden Markov Model (CSPHMM3)" as a classifier. An Emirati-accented (Arabic) speech database with "Mel-Frequency Cepstral Coefficients" as the extracted features has been used to evaluate our work. Our results demonstrate that speaker verification accuracy based on CSPHMM3 is greater than that based on the "state-of-the-art classifiers and models such as Gaussian Mixture Model (GMM), Support Vector Machine (SVM), and Vector Quantization (VQ)".

## I. INTRODUCTION AND LITERATURE REVIEW

Automatic speaker recognition (ASR) is subdivided into two parts: Automatic Speaker Verification (ASV) and Automatic Speaker Identification (ASI). ASV is defined as admitting or denying the claimed speaker from his/her voice. On the other hand, ASI is termed as identifying the undetermined speaker using his/her voice from a set of known speakers. "ASV technologies have ample span of usages such as: biometric person verification, SV for surveillance, forensic speaker recognition, and security applications comprising of credit card transactions and computer access control. ASI can be used in inspecting criminals to decide who uttered the voice collected during the misdeed [1]. ASR is classed, based on the spoken text, into text-dependent and text-independent types. In the text-dependent type, ASR requires the speaker to utter speech of the same text in both training and testing phases; on the other hand, the text-independent type, ASR is independent on the text being spoken".

Speaker recognition has been studied recently using Arabic speech database in each of neutral [2], [3], [4] and emotional/stressful talking environments [5], [6], [7].

In neutral talking environment, Shahin focused in one of his research [2] on testing a "text-independent speaker verification using Emirati-accented speech corpus captured in a such environment. Twenty five Emirati native speakers per gender uttered eight commonly-used Emirati sentences. Mel-Frequency Cepstral Coefficients (MFCCs) have been utilized as the extracted features of speech signals. Three various classifiers have been employed in his work. These classifiers are: First-Order Hidden Markov Models (HMM1s), Second-Order Hidden Markov Models (HMM2s), and Third-Order Hidden Markov Models (HMM3s). He concluded that HMM3s perform better than each of HMM1s and HMM2s for a text-independent Emirati-accented speaker verification in neutral environment. Alsulaiman *et al.* [3] researched Arabic speaker recognition utilizing an openly available speech dataset named Babylon Levantine that is accessible from the Linguistic Data Consortium (LDC). In their work, they used Hidden Markov Models (HMMs) as classifiers. Their outcomes presented that the recognition accuracy improves as the number of mixtures increases, until it comes to a saturation point which depends on the number of HMM states and the data size. Alarifi *et al.* [4] proposed a modern Arabic text-dependent speaker verification system for mobile devices using Artificial Neural Networks (ANNs)" to perform verification on the authorized user and unlock his/her machine.

In emotional/stressful talking environments, Shahin *et al.* [5] enhanced "text-independent speaker identification accuracy in emotional talking environments based on novel classifier named cascaded Gaussian Mixture Model-Deep Neural Network (GMM-DNN). Their work focused on proposing, applying, and testing a new framework for speaker identification in such talking environments based on sequential Gaussian Mixture Model followed by Deep Neural Network as a classifier. Their findings showed that the cascaded GMM-DNN classifier enhances speaker identification accuracy at different emotions using two diverse speech datasets: Emirati Arabic speech database and Speech Under Simulated and Actual Stress (SUSAS) English dataset. Their proposed classifier outperforms traditional classifiers such as Multilayer Perceptron (MLP) and Support Vector Machine (SVM) for the two databases. In another work by Shahin *et al.* [6], where their contribution was devoted to capturing Emirati-accented speech database in each of neutral and shouted talking environments to investigate and improve text-independent Emirati-accented speaker identification performance in shouted environment based on each of First-Order Circular Suprasegmental Hidden Markov Models (CSPHMM1s), Second-Order Circular Suprasegmental Hidden Markov Models (CSPHMM2s), and Third-Order Circular Suprasegmental Hidden Markov Models (CSPHMM3s) as classifiers. In their work, the database was gathered from fifty Emirati native speakers (twenty five per gender) talking eight popular Emirati sentences in each of neutral and shouted talking environments. The extracted features of their gathered database are MFCCs. Their results showed that average Emirati-accented speaker identification accuracy in neutral environment is 94.0%, 95.2%, and 95.9% based on CSPHMM1s, CSPHMM2s, and CSPHMM3s, respectively. On the other side, the average accuracy in shouted environment is 51.3%, 55.5%, and 59.3% based, respectively, on CSPHMM1s, CSPHMM2s, and CSPHMM3s. In one more study by Shahin [7], he spotlighted his work on enhancing text-independent Emirati-accented speaker identification accuracy in emotional environments based on the classifiers: HMM1s, HMM2s, and HMM3s. In his work, the corpus was collected from twenty five Emirati native speakers per gender speaking eight frequent Emirati sentences in each of neutral, angry, sad, happy, disgust, and fear emotions. The extracted features of his collected database are MFCCs. His results showed that average Emirati-accented speaker identification accuracy in emotional environments is 58.8%, 61.8%, and 65.9% based on HMM1s, HMM2s, and HMM3s, respectively.

This work focuses on studying and enhancing speaker verification accuracy in emotional talking environments using Emirati-accented corpus based on CSPHMM3 as a classifier. In addition, some additional experiments have been executed in this work to thoroughly test the attained outcomes.

The rest of this paper is given as: The basics of CSPHMM3 are presented in Section II. The utilized dataset and extraction of features appear in Section III. The algorithm of speaker verification based on CSPHMM3 and the experiments are explained in Section IV. Decision threshold is covered in Section V. The attained findings and the experiments are discussed in Section VI. Concluding remarks of this study are considered in Section VII".

## II. BASICS OF CSPHMM3

"Third-Order Circular Suprasegmental Hidden Markov Model has been developed from acoustic Third-Order Hidden Markov Model (HMM3) [8]. Shahin [9] proposed, employed, and valued HMM3 to alleviate the declined text-independent speaker identification accuracy in a shouted talking environment".

### A. Basics of HMM3

In "HMM1, the underlying state sequence is a first-order Markov chain where the stochastic process is specified by a 2-D matrix of a priori transition probabilities ($a_{ij}$) between states $s_i$ and $s_j$ where $a_{ij}$ is given as [10],

$$a_{ij} = \text{Prob}\left(q_t = s_j | q_{t-1} = s_i\right) \quad (1)$$

In HMM2, the underlying state sequence is a second-order Markov chain where the stochastic process is described by a 3-D matrix ($a_{ijk}$). Hence, the transition probabilities in HMM2 are given as [11],

$$a_{ijk} = \text{Prob}\left(q_t = s_k | q_{t-1} = s_j, q_{t-2} = s_i\right) \quad (2)$$

with the constraints,

$$\sum_{k=1}^{N} a_{ijk} = 1 \quad N \geq i, j \geq 1$$

In HMM3, the underlying state sequence is a third-order Markov chain where the stochastic process is stated by a 4-D matrix ($a_{ijkw}$). Subsequently, the transition probabilities in HMM3 are given as [9],

$$a_{ijkw} = \text{Prob}\left(q_t = s_w | q_{t-1} = s_k, q_{t-2} = s_j, q_{t-3} = s_i\right) \quad (3)$$

with the constraints,

$$\sum_{w=1}^{N} a_{ijkw} = 1 \quad N \geq i, j, k \geq 1$$

The probability of the state sequence, $Q \triangleq q_1, q_2, ..., q_T$, is expressed as:

$$\text{Prob}(Q) = \Psi_{q_1} a_{q_1 q_2 q_3} \prod_{t=4}^{T} a_{q_{t-3} q_{t-2} q_{t-1} q_t} \quad (4)$$

where $\Psi_i$ is the probability of a state $s_i$ at time $t = 1$, $a_{ijk}$ is the probability of the transition from a state $s_i$ to a state $s_k$ at time $t = 3$. $a_{ijk}$ can be computed from equation (2). Thus, the initial parameters of HMM3 can be obtained from the trained HMM2.

Given a sequence of observed vectors, $O \triangleq O_1, O_2, ..., O_T$, the joint state-output probability is expressed as [9]:

$$\text{Prob}(Q, O | \lambda) = \Psi_{q_1} b_{q_1}(O_1) a_{q_1 q_2 q_3} b_{q_3}(O_3) \prod_{t=4}^{T} a_{q_{t-3} q_{t-2} q_{t-1} q_t} b_{q_t}(O_t) \quad (5)$$

Readers can find further explanations about these three models from [9], [10], [11].

### B. CSPHMM3

Within Third-Order Circular Hidden Markov Model (CHMM3), prosodic and acoustic information can be merged into CSPHMM3" as given by the formula [12],

$$\log P\left(\lambda^v_{\text{CHMM3s}}, \Psi^v_{\text{CSPHMM3s}} | O\right) = (1-\alpha) \cdot \log P\left(\lambda^v_{\text{CHMM3s}} | O\right) \quad (6)$$
$$+ \alpha \cdot \log P\left(\Psi^v_{\text{CSPHMM3s}} | O\right)$$

where "$\lambda^v_{\text{CHMM3}}$ is the acoustic third-order circular hidden Markov model of the $v^{\text{th}}$ speaker and $\Psi^v_{\text{CSPHMM3}}$ is the suprasegmental third-order circular hidden Markov model of the $v^{\text{th}}$ speaker". Figure 1 presents an example of a fundamental structure of CSPHMM3 that has been developed from CHMM3. This figure contains of "six third-order acoustic hidden Markov states: $q_1, q_2, ..., q_6$ positioned in a circular form. $p_1$ is a third-order suprasegmental state that is made up of $q_1$, $q_2$, and $q_3$. $p_2$ is a third-order suprasegmental state which is composed of $q_4$, $q_5$, and $q_6$". "The suprasegmental states $p_1$ and $p_2$ are located in a circular form. $p_3$ is a third-order suprasegmental state that is comprised of $p_1$ and $p_2$".

## III. SPEECH DATASET AND EXTRACTION OF FEATURES

### A. Speech Dataset

In this study, our work has been experimented on an "Emirati-accented Arabic speech dataset captured from 15 male and 15 female local Emirati speakers. These speakers utter 8 familiar Emirati sentences that are frequently spoken in the UAE society. Each sentence has been spoken by each speaker 9 times under each of neutral, happy, sad, disgust, angry, and fear emotions. Table I displays the database used in this research where the right column shows the utterances in Emirati accent, the left column displays the English version, and the middle column explains the phonetic transcriptions of these utterances. This corpus was collected in two isolated and diverse sessions: training session and testing session. The dataset was recorded in a clean environment in the College of Communication, University of Sharjah, United Arab Emirates by a set of specialized engineering students. The dataset was collected by a speech acquisition board using a 16-bit linear coding A/D converter and sampled at a sampling rate of 44.6 kHz".

### B. Extraction of Features

In the present work, the "phonetic content of speech signals" in our corpus is described by "Mel-Frequency Cepstral Coefficients (static MFCCs) and delta Mel-Frequency Cepstral Coefficients (delta MFCCs)". Such coefficients have been mostly used in many studies in the fields of "speaker recognition [13], [14], [15] and emotion recognition" [16], [17], [18], [19]. In this study, "MFCC feature analysis" is utilized to create the "observation vectors in each of CHMM3 and CSPHMM3".

A 32-dimension "MFCC (16 static MFCCs and 16 delta MFCCs) feature analysis" is used to produce the "observation vectors" in every model of "CHMM3 and CSPHMM3". The "number of conventional states, $N$, in every model is 6 and the number of suprasegmental states is two (each suprasegmental state is made up of three conventional states) in CSPHMM3 with a continuous mixture observation density has been chosen for each model".

## IV. SPEAKER VERIFICATION ALGORITHM BASED ON CSPHMM3 AND THE EXPERIMENTS

The "training phase of CSPHMM3 is highly identical to the training phase of CHMM3. In the training phase of CSPHMM3, suprasegmental third-order circular model is trained on top of acoustic third-order circular model. In the training phase of CSPHMM3s, the $v$th speaker has been denoted by a $v$th model. The $v$th model has been obtained using the first four sentences with a duplication of nine utterances per sentence of the corpus. This provides a sum of 36 utterances (4 sentences × 9 repetitions) for each speaker model.

In the test phase of CSPHMM3, every one of the thirty speakers utilized nine utterances for each sentence of the last four sentences (text-independent) of the corpus. The entire number of utterances used in this phase is 1080 (30 speakers × 4 sentences × 9 utterances/sentence). In this study, 12 speakers for every gender have been used as claimants and the remaining of the speakers have been used as imposters".

The "log-likelihood ratio" in the log domain has been performed to authenticate the speaker identity based on CSPHMM3s as given in the following formula [20],

$$\Lambda_{CSPHMM3s}(O) = \log\left[P\left(O|\lambda_{CHMM3s,C}, \Psi_{CSPHMM3s,C}\right)\right] \\ - \log\left[P\left(O|\lambda_{CHMM3s,\overline{C}}, \Psi_{CSPHMM3s,\overline{C}}\right)\right] \quad (7)$$

where, $\Lambda_{CSPHMM3s}(O)$ is the "log-likelihood ratio in the $log$ domain", $P\left(O|\lambda_{CHMM3s,C}, \Psi_{CSPHMM3s,C}\right)$ is the probability of the observation sequence $O$ given it arises from the claimed speaker, and $P\left(O|\lambda_{CHMM3s,\overline{C}}, \Psi_{CSPHMM3s,\overline{C}}\right)$ is the probability of the observation sequence $O$ given it does not come from the claimed speaker.

The probability of the observation sequence $O$ provided it occurs from the claimed speaker can be calculated as [20],

$$\log P\left(O|\lambda_{CHMM3s,C}, \Psi_{CSPHMM3s,C}\right) = \frac{1}{T}\sum_{t=1}^{T}\log P\left(o_t|\lambda_{CHMM3s,C}, \Psi_{CSPHMM3s,C}\right) \quad (8)$$

where, $O = o_1 o_2 ... o_t ... o_T$ and $T$ is the utterance duration.

The probability of the observation sequence $O$ given it does not arise from the claimed speaker can be calculated using a set of $B$ imposter speaker models: $\{\lambda_{CHMM3s,\overline{C}_1}, \Psi_{CSPHMM3s,\overline{C}_1},...,\lambda_{CHMM3s,\overline{C}_B}, \Psi_{CSPHMM3s,\overline{C}_B}\}$ as,

$$\log P\left(O|\lambda_{CHMM3s,\overline{C}}, \Psi_{CSPHMM3s,\overline{C}}\right) = \left\{\frac{1}{B}\sum_{b=1}^{B}\log\left[P\left(O|\lambda_{CHMM3s,\overline{C}_b}, \Psi_{CSPHMM3s,\overline{C}_b}\right)\right]\right\} \quad (9)$$

where $P\left(O|\lambda_{CHMM3s,\overline{C}_b}, \Psi_{CSPHMM3s,\overline{C}_b}\right)$ can be computed using Eq. (8).

## V. DECISION THRESHOLD

There are two classes of error that can occur in speaker verification problem. The two classes are "false rejection and false acceptance". When a valid identity claim is denied, it is termed a "false rejection error"; on the other scale, when the identity claim from an imposter is permitted, it is entitled a "false acceptance error".

Speaker verification problem demands building a binary decision based on two hypotheses: "Hypothesis $H_0$ if the observation sequence $O$ given it comes from the claimed speaker or hypothesis $H_1$ if the observation sequence $O$ given it does not come from the claimed speaker.

To permit or deny the claimed speaker, a comparison between the log-likelihood ratio and the threshold (θ) should be made as the last stage in the authentication practice, i.e., [20]:

Admit the claimed speaker if $\Lambda(O) \geq \theta$

Refuse the claimed speaker if $\Lambda(O) < \theta$

Open set speaker verification usually uses thresholding to decide if a speaker is out of the set. Both kinds of error in speaker verification problem rely on the threshold utilized in the decision making practice. A firm value of threshold makes it difficult for false speakers to be incorrectly accepted but at the cost of mistakenly denying true speakers. On the other side, an eased value of threshold makes correct speakers to be always permitted at the expenditure of wrongly admitting incorrect speakers. To assign a fitting value of threshold that agrees with a needed level of a true speaker refusal and a false speaker approval, it is essential to know the distribution of true speaker and false speaker scores. An adequate procedure for placing a value of threshold is to begin with a relaxed starting value of threshold and then let it regulate by putting it to the average of recent trial scores. This relaxed value of threshold causes deficient shield against wrong speaker tryouts".

## VI. RESULTS AND DISCUSSION

Table II illustrates percentage Equal Error Rate (EER) for speaker verification in emotional environments based on CHMM3 and CSPHMM3. The average value of percentage EER is 29% and 21.8% based on CHMM3 and CSPHMM3, respectively. This table illustrates that the minimum percentage EER happens when speakers speak neutrally, while the maximum percentage EER occurs when speakers talk angrily. This table clearly gives higher percentage EER when speakers speak emotionally compared to when speakers speak neutrally. It is also apparent from this table that CSPHMM3 performs better than CHMM3 for speaker verification in emotional environments.

To confirm whether EER differences (EER based on CSPHMM3s and that based on CHMM3s) are tangible or simply originate from statistical differences, a statistical significance test has been conducted. The statistical significance test has been applied based on the "Student's $t$ Distribution test" as given by,

$$t_{model\ x,\ model\ y} = \frac{\overline{X}_{model\ x} - \overline{X}_{model\ y}}{SD_{pooled}} \quad (10)$$

where $\overline{X}_{model\ x}$ is the mean of the first sample (model $x$) of size $n$, $\overline{X}_{model\ y}$ is the mean of the second sample (model $y$) of the same size, and $SD_{pooled}$ is the pooled standard deviation of the two samples (models $x$ and $y$) given as,

$$SD_{pooled} = \sqrt{\frac{SD_{model\ x}^2 + SD_{model\ y}^2}{2}} \quad (11)$$

where $SD_{model\ x}$: is an estimate of the standard deviation of the average of the first sample (model $x$) of size $n$ and $SD_{model\ y}$ is an estimate of the standard deviation of the average of the second sample (model $y$) of the same size.

The calculated $t$ value between CSPHMM3 and CHMM3 is computed based on Table II. The computed value is $t_{CSPHMM3,\ CHMM3} = 1.919$ which is bigger than the tabulated critical value $t_{0.05} = 1.645$ at 0.05 significant level. Thus, it is obvious that CSPHMM3 outperforms CHMM3 for "speaker verification in emotional environments".

Speaker verification accuracy based on "CSPHMM3 has been compared with that based on the state-of-the-art classifiers and models such as Gaussian Mixture Model (GMM) [20], Support Vector Machine (SVM) [21], and Vector Quantization (VQ)" [22]. The average "EER for speaker verification system in emotional environments based on GMM, SVM, and VQ" yields 26.3%, 24.6%, and 25.8%, respectively. It is evident from this experiment that CSPHMM3 produces less EER than GMM, SVM, and VQ by 20.6%, 12.8%, and 18.3%, respectively.

An "informal subjective assessment for speaker verification using our collected corpus has been performed using 10 human non-professional adult listeners. In this assessment, a sum of 540 utterances (30 speakers × 6 emotions × 3 repetitions) have been utilized. These listeners are questioned to verify the unknown speakers. Based on this assessment, the average EER using our collected corpus is 22.1%. This average EER is similar to the attained average based on CSPHMM3 (21.8%)".

## VII. CONCLUDING REMARKS

In this research, "CHMM3 and CSPHMM3" have been used as classifiers to verify the claimed speaker who is talking emotionally in "Emirati Arabic language". Some concluding remarks can be drawn in this study. First, "CSPHMM3" is superior to each of "CHMM3, GMM, SVM, and VQ for speaker verification in emotional talking environments". Second, the highest speaker verification accuracy occurs when speakers speak neutrally. Finally, the lowest speaker verification accuracy happens when speakers speak angrily.

This work has some restrictions. Firstly, our dataset is restricted to thirty speakers only. Secondly, the attained speaker verification accuracy based on CSPHMM3 is not ideal. Our future plan is to apply Deep Neural Network (DNN) to achieve better results [23]. Also, our coming plan is to study and investigate Emirati-accented speaker verification in biased emotional talking environments [24], [25].

## ACKNOWLEDGMENT

The authors of this study wish to convey their appreciations to the "University of Sharjah" for supporting this work through the competitive research project entitled "Capturing, Studying, and Analyzing Arabic Emirati-Accented Speech Database in Stressful and Emotional Talking Environments for Different Applications, No. 1602040349-P".

Table I. Emirati-accented speech dataset in its: English version, phonetic transcriptions, and Emirati accent

| No. | English Version | Phonetic Transcriptions | Emirati Accent |
|---|---|---|---|
| 1. | We will meet with you in an hour. | / bɪntlɑːga wɪjɑːk ʕugub sɑːʕah / | بنتلاقى وياك عقب ساعة |
| 2. | Go to my father he wants you. | /siːr ʕɪnd abuːjeh yibɑːk / | سير عند ابويه يباك |
| 3. | Bring my cell phone from the room. | /hɑːt tɪlɪfuːniː mɪnɪl ḥɪjrah / | هات تيلفوني من الحجرة |
| 4. | I am busy now I will talk to you later. | / maʃɣɔːɫ(a) ʌḥiːn baramsɪk ʕʌb sɑːʕəh / | مشغول/مشغولة الحين برمسك عقب |
| 5. | Every seller praises his market. | / kɪl byaɪʕ yɪmdeḥ suːgah / | كل بياع يمدح سوقه |
| 6. | A stranger is a wolf whose bite wounds won't heal. | / ɪlɣariːb ðiːb w ʕaẓitah maṭiːb / | الغريب ذيب و عضته ما تطيب |
| 7. | Show respect around some people and show self-respect around other people. | / naːæs ɪḥʃɪmhom w naːs ɪḥʃɪm nafsak ʕanhom / | ناس احشمهم و ناس احشم نفسك عنهم |
| 8. | Don't criticize what you can't get and don't swirl around something you can't obtain. | / illi magdart tɪyibah lɑː tʕiːbah w illi mɑːṭuːlah lɑː ṭhuːm ḥuːlah / | اللي ما قدرت تييبه لا تعيبه و اللي ما تطوله لا تحوم حوله |

Table II. EER using Emirati-Accented database based on "CHMM3 and CSPHMM3"

| Model | Gender | "Speaker recognition accuracy under each emotion" (%) | | | | | |
|---|---|---|---|---|---|---|---|
| | | "Neutral" | "Happiness" | "Sadness" | "Disgust" | "Anger" | "Fear" |
| CHMM3 | Male | 6 | 30 | 34 | 35 | 38 | 32 |
| | Female | 5 | 29 | 32 | 36 | 37 | 34 |
| | Average | 5.5 | 29.5 | 33 | 35.5 | 37.5 | 33 |
| CSPHMM3 | Male | 4 | 23 | 24 | 26 | 29 | 25 |
| | Female | 4 | 22 | 24 | 26 | 29 | 25 |
| | Average | 4 | 22.5 | 24 | 26 | 29 | 25 |

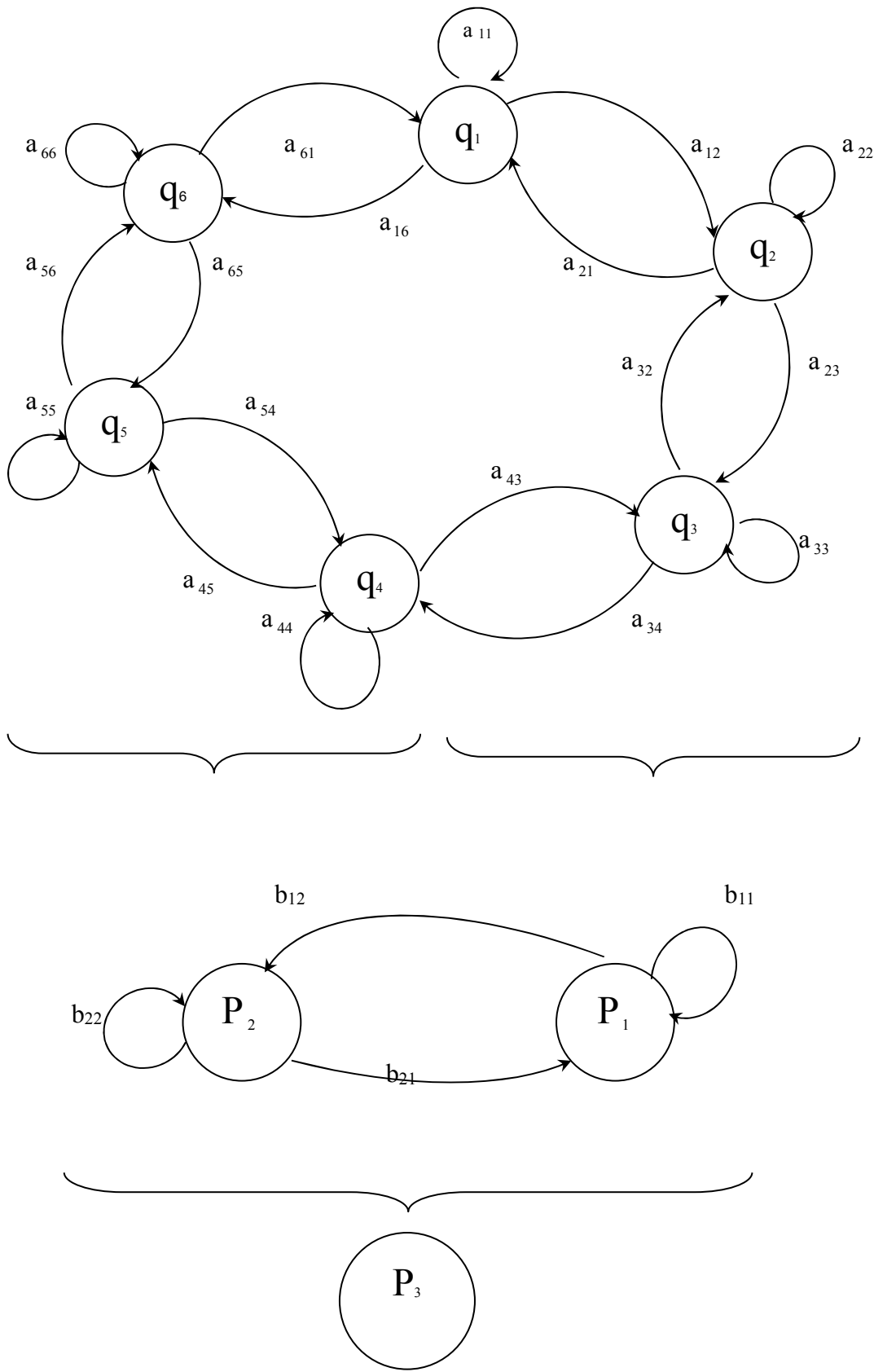

Fig. 1. Basic structure of CSPHMM3 developed from CHMM3